\documentclass{llncs}
\usepackage[T1]{fontenc}
\usepackage[utf8]{inputenc}
\usepackage[english]{babel}
\usepackage{tikz}
\usepackage{amsmath}            
\usepackage[font=small,skip=3pt]{caption}
\usetikzlibrary{quantikz}
\usetikzlibrary{shapes.geometric,positioning,calc,arrows.meta}
\usepackage{graphicx}
\usepackage{amsfonts}
\usepackage{comment}
\usepackage{algorithm}
\usepackage{algorithmic}
\usepackage{xcolor}
\usepackage{enumitem}
\usepackage{booktabs}
\usepackage{float}      
\usepackage{placeins}   
\usepackage{newunicodechar}
\usepackage[hidelinks]{hyperref}

\newunicodechar{≈}{\approx}

\usepackage[font=small,skip=3pt]{caption}

\usepackage{etoolbox}
\AtBeginEnvironment{tabular}{\scriptsize} 
\renewcommand{\arraystretch}{0.95}         

\setlist{leftmargin=*, itemsep=1pt, topsep=2pt, parsep=0pt, partopsep=0pt}

\begin{document}
\title{Physics-Guided Linear Mapper for Quantum Error Mitigation}
%
%

\author{Tulsi Chaudhari\inst{1} \and Krishna Bhatia\inst{2} \and Shalini Devendrababu\inst{2} \and Srinjoy Ganguly\inst{3} \and Luis Gerardo Ayala Bertel\inst{4}}

\institute{%
Florida International University, USA \and
Quantum AI Lab, Fractal AI Research, India \and
University College London, Gower Street, London, UK \and
Cartagena University, Colombia, South America \newline
tchaudhari049@gmail.com
krishna.bhatia@fractal.ai
shalini.devendrababu@fractal.ai
layalab@unicartagena.edu.co
}

\maketitle

%

\begin{abstract}
We introduce a novel physics-guided linear mapper (PGLM) for quantum error mitigation that uses seven distinct interpretable features derived from circuit complexity and device calibration data. The goal is to provide a data-efficient, interpretable, and low-latency alternative to the black-box machine learning for quantum error mitigation in noisy-intermediate scale quantum devices. Evaluated on 52 simulated benchmark circuits (1--4 qubits), PGLM demonstrates strong performance in noise-accumulation regimes: 50.1\% RMSE reduction on 3-qubit circuits and 32.3\% on 4-qubit circuits, while single-qubit circuits show degraded performance. A circuit-size-aware deployment policy achieves 32.6\% aggregate improvement. Sub-millisecond inference enables integration into variational algorithms, and analysis of learned coefficients reveals that circuit depth and CNOT count dominate error prediction, consistent with decoherence mechanisms. Results are simulator-based with idealized noise models; hardware validation remains essential future work.
\end{abstract}


%
%
\section{Introduction}

Quantum error mitigation (QEM) has emerged as a critical enabling technology for near-term quantum computing applications, offering a path to extract meaningful results from noisy intermediate-scale quantum (NISQ) devices without the overhead of full quantum error correction. While traditional QEM methods such as zero-noise extrapolation (ZNE) and probabilistic error cancellation (PEC) have demonstrated theoretical efficacy~\cite{Temme2017,Kandala2019}, their practical deployment is often hindered by exponential sampling overheads that scale poorly with circuit complexity~\cite{Qin2023}.

\vspace{2mm}

Recent advances in machine learning for quantum error mitigation (ML-QEM)~\cite{Strikis2021,Liao2023} promise to address these scalability challenges by learning statistical relationships between noisy and ideal quantum computation outcomes.Recent advances in machine learning for quantum error mitigation (ML-QEM)~\cite{Strikis2021,Liao2023} promise to address these scalability challenges by learning statistical relationships between noisy and ideal quantum computation outcomes. However, prior ML-QEM studies~\cite{Strikis2021,Liao2023,Adeniyi2025,GTranQEM2025} say that existing approaches typically require extensive training datasets that are expensive to generate on quantum hardware, and they rely on black-box models that provide little physical insight into the error mitigation process.

\vspace{2mm}

This work introduces the Physics-Guided Linear Mapper (PGLM), a novel approach to ML-QEM that addresses both limitations through a carefully designed feature space incorporating physical characteristics of quantum noise accumulation. Rather than treating error mitigation as generic regression, PGLM leverages domain knowledge about how quantum errors scale with circuit properties such as gate count, depth, and device calibration parameters.

\vspace{2mm}

Our approach makes three key contributions:

\begin{enumerate}[label=\arabic*., leftmargin=*]
  \item PGLM achieves effective error mitigation using only 50--200 training circuits, representing an order-of-magnitude reduction compared to typical neural network approaches~\cite{Liao2023,Adeniyi2025}. This data efficiency stems from physics-informed basis functions encoding prior knowledge about noise behavior. \newline

  \item Unlike black-box ML models~\cite{Strikis2021,GTranQEM2025}, PGLM's linear architecture with physics-guided features provides interpretable coefficients revealing how different error sources contribute to expectation value corrections, enabling both scientific understanding and practical diagnostics. \newline

  \item Through systematic evaluation across circuit sizes, we demonstrate that PGLM's effectiveness correlates with the physical regime where noise accumulation effects dominate, specifically multi-qubit circuits where our physics-guided approach achieves up to 50\% RMSE reduction over raw noisy results. \newline

\end{enumerate}

In this paper, we address a practical gap in ML-QEM: methods that are both data-efficient and physically interpretable while incurring minimal latency for on-line use. Concretely, we propose the Physics-Guided Linear Mapper (PGLM), a seven-feature ridge regressor designed to capture dominant noise-accumulation mechanisms with order-of-magnitude lower training data than neural baselines. The remainder of the paper is organised as follows. Section 2 reviews background and related work and highlights how PGLM differs from prior linear and physics-informed approaches. Section 3 presents the PGLM model and feature design. Section 4 describes experimental setup and noise models. Section 5 reports results and ablations. Section 6 discusses practical deployment, limitations, and future directions. We conclude in Section 7.

\section{Background and Related Work}

Quantum circuits in NISQ devices suffer from several fundamental error mechanisms~\cite{Temme2017,Kim2023}: decoherence (coherence decay with time constant $T_2$), gate errors (imperfect unitary implementation with fidelities $\sim$98-99.8\%), measurement errors (readout bit-flips with probability $\sim$1-3\%), thermal relaxation (excited state decay with time constant $T_1$), and dephasing (phase randomization). These errors compound across circuit depth, motivating error mitigation strategies.

\subsection{Traditional Quantum Error Mitigation Methods}
Quantum error mitigation encompasses a family of techniques designed to suppress the effects of noise in quantum computations without requiring the full overhead of quantum error correction. These methods sacrifice the guarantee of fault-tolerant computation in exchange for practical applicability to near-term quantum devices with limited coherence times and gate fidelities.

\paragraph{Zero-Noise Extrapolation (ZNE):}
ZNE is based on artificially amplifying noise to extrapolate results to the zero-noise limit~\cite{Temme2017,GiurgicaTiron2020,He2020_ZNE}. The method involves: (1) noise scaling (e.g., unitary folding \(U \rightarrow U U^{\dagger} U\)); (2) execution at multiple noise levels; and (3) extrapolation of expectation values to zero noise via polynomial or exponential fitting. Practical deployment faces multiplicative sampling overheads (often 10--100\(\times\))~\cite{Kandala2019} and increased depth that exacerbates decoherence.

\paragraph{Probabilistic Error Cancellation (PEC):}
PEC decomposes the ideal channel as a linear combination of implementable noisy channels: \(\mathcal{E}^{-1} = \sum_i \alpha_i \mathcal{F}_i\)~\cite{Temme2017,Song2019}. Quasi-probability sampling induces an exponential sampling overhead scaling as \(\gamma^G\) for \(G\) gates with normalization \(\gamma>1\), limiting practicality beyond modest circuit sizes.

\paragraph{Clifford Data Regression (CDR):}
CDR trains on efficiently simulable Clifford circuits to predict ideal outcomes for non-Clifford circuits~\cite{Czarnik2020,Czarnik2025}. While often polynomial in overhead and effective when non-Clifford gates are sparse, it requires hundreds to thousands of training evaluations and depends on similarity between training and target circuits.

\paragraph{Fundamental Limitations:}
No-go results show sampling complexity for \(\epsilon\)-accurate mitigation can scale exponentially with depth in the worst case~\cite{Qin2023}, motivating targeted, structure-exploiting approaches.

\subsection{Recent ML-QEM Approaches and Data Requirements}
Neural models (e.g., random forests, MLPs, GNNs) can learn mappings from noisy to ideal expectation values~\cite{Strikis2021,Adeniyi2025,GTranQEM2025}, but generally need thousands of training circuits and substantial classical training time. Adaptive/context-aware methods~\cite{Jiang2024,Liao2023} achieve strong accuracies yet heighten data demands. Physics-inspired architectures~\cite{PhysicsInspired2025,Babukhin2023} (e.g., neural noise-accumulation surrogates) reduce data by \(\sim\)an order of magnitude but still lack direct interpretability and require sophisticated training. Overall, data hunger, training overhead, and black-box behavior hinder deployment.

\subsection{Physics-Informed Machine Learning in Quantum Computing}
Let a circuit \(\mathcal{C}\) implement \(\mathcal{E}=\mathcal{N}\circ\mathcal{U}\). For observable \(O\),
\begin{equation}
\langle O\rangle_{\text{noisy}}=\mathrm{Tr}\!\big(O\,\mathcal{N}(\mathcal{U}(\rho_0))\big)
\label{eq:noisy_expectation}
\end{equation}

Physics suggests error accumulation scales with structure~\cite{Temme2017}: for small gate error \(p_{\text{gate}}\),
\begin{equation}
p_{\text{error}} \approx 1-(1-p_{\text{gate}})^G \approx G\,p_{\text{gate}}
\label{eq:p_error_scaling}
\end{equation}

Under depolarizing noise with parameter \(p\),
\begin{equation}
\langle O\rangle_{\text{noisy}}=(1-p)^{n_{\text{gates}}}\langle O\rangle_{\text{ideal}}+\frac{p}{2^n}\mathrm{Tr}(O)
\label{eq:depolarizing_relation}
\end{equation}

Hence multiplicative features such as \(\langle O\rangle_{\text{noisy}}\times f(\text{circuit})\) are natural. Device parameters (e.g., \(T_1,T_2\), readout error) can be encoded via physics-motivated basis functions~\cite{Bravyi2021_readout,NguyenReadout2023}, e.g.
\begin{equation}
\phi_{\text{coh}}=\langle O\rangle_{\text{noisy}}\exp\!\left(-\frac{T_{\text{circuit}}}{T_2}\right),\quad
\phi_{\text{ro}}=\langle O\rangle_{\text{noisy}}\,(1-2p_{\text{readout}})
\label{eq:feature_examples}
\end{equation}

In linear models \begin{equation}
\hat{y}=\sum_i w_i\phi_i
\label{eq:linear_model_general}
\end{equation}
, weights map to mechanisms, enabling interpretability and principled transfer.

As summarised in Table~\ref{tab:literature}, current methods trade off interpretability, data efficiency, and deployability. Neural models may outperform traditional QEM but lack transparency and require \(10^3\!-\!10^4\) circuits. Practical workflows also impose latency constraints (\(T_{\text{mitigation}} + T_{\text{classical}} < T_{\text{coherence}} - T_{\text{execution}}\)). A missing paradigm is one that is data-efficient (\(N_{\text{train}}\ll10^3\)), interpretable, fast at inference, transferable across devices, and theoretically grounded---motivating physics-guided approaches like PGLM.

\begin{table}[t]
\centering
\caption{Comparison of quantum error mitigation (QEM) and ML-QEM methods.}
\label{tab:literature}
\scriptsize
\setlength{\tabcolsep}{3pt}
\renewcommand{\arraystretch}{1.2}
\begin{tabular}{|p{1.7cm}|p{2.8cm}|p{1.9cm}|p{1.8cm}|p{1.3cm}|p{2.8cm}|}
\hline
\textbf{Approach} & \textbf{Core idea} & \textbf{Quantum overhead} & \textbf{Train data} & \textbf{Interpre- tability} & \textbf{Key limits} \\
\hline
ZNE~\cite{Temme2017,Kandala2019} & Scale noise, extrapolate to zero-noise & $10$--$100\times$ runs & none & High & Deeper circuits, unstable fits \\
\hline
PEC~\cite{Temme2017,Song2019} & Quasi-probability inversion of noisy channels & Exponential in gates & none & Medium & Exponential sampling overhead \\
\hline
CDR~\cite{Czarnik2020,Czarnik2025} & Train on Clifford circuits, predict non-Clifford & Polynomial & $10^2$--$10^3$ & Medium & Large training need; limited transfer \\
\hline
NN-ML~\cite{Strikis2021,Liao2023} & Black-box regressors (MLP, RF, GNN) & Low & $10^3$--$10^4$ & Low & Data hungry; opaque \\
\hline
Phys.-insp.\ NN~\cite{PhysicsInspired2025,Babukhin2023} & Encode noise physics in NN & Low & $10^2$--$10^3$ & Low--Med. & Training heavy; less interpretable \\
\hline
\textbf{PGLM (ours)} & Physics-guided linear features & \textbf{Low} & \textbf{50--200} & \textbf{High} & Linear limits; weak in 1-qubit regime \\
\hline
\end{tabular}
\end{table}

\paragraph{Distinguishing PGLM from prior linear/feature-driven QEM}
\paragraph{Distinguishing PGLM from prior linear/feature-driven QEM}
Several prior works have explored simple regression or physics-inspired neural networks for QEM. PGLM differs on three axes:
\begin{enumerate}[label=\arabic*., leftmargin=*]
  \item \textbf{Feature set.} We use a minimal, physically motivated 7-dimensional basis that mixes multiplicative and additive terms (e.g., $y_{\text{noisy}}\times n_{\text{cx}}$, $y_{\text{noisy}}\times d$) selected to reflect gate-dominated versus decoherence-dominated error channels.
  \item \textbf{Deployment focus.} We optimise for sub-millisecond inference and sub-second training suitable for on-device classical controllers, so the model can be used in real-time workflows.
  \item \textbf{Interpretability.} By using ridge regression with fixed standardisation and a stability-oriented regularisation ($\alpha=1$), the learned coefficients map directly to physical mechanisms, enabling principled device transfer and simple adaptive scaling. We include detailed ablations (Sec.~5) that isolate the benefits of each design choice and compare against representative baselines.
\end{enumerate}

\section{Methodology}
\label{sec:method}



\begin{figure}[H]
  \centering
  \includegraphics[width=\linewidth]{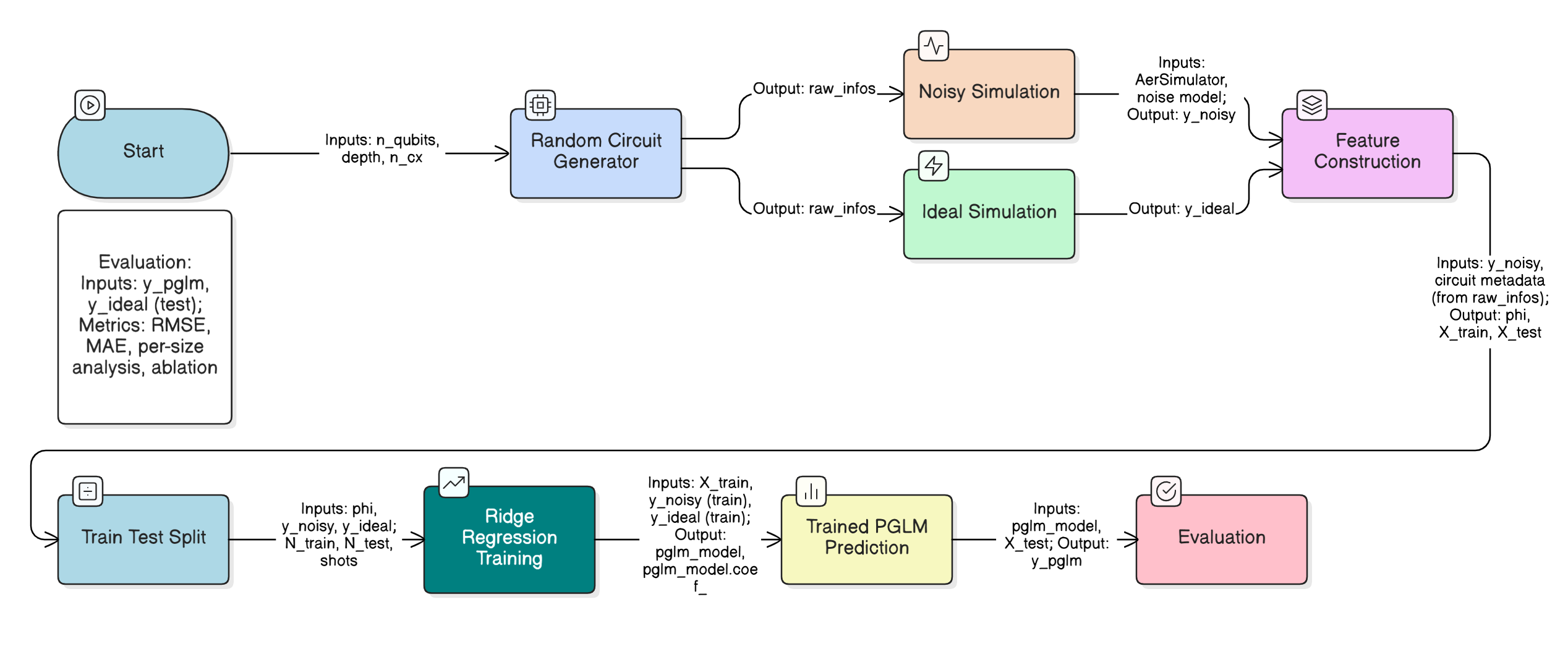}
  \caption{PGLM workflow: random circuit generation, noisy/ideal simulation, physics-guided feature construction, ridge training, prediction, and evaluation.}
  \label{fig:qem_pipeline}
\end{figure}
\FloatBarrier

\subsection{Problem Setup and Physics-Guided Feature Engineering}
\label{subsec:features}
The overall PGLM workflow is illustrated in Fig.~\ref{fig:qem_pipeline}. We observe a noisy expectation $y_{\text{noisy}}=\mathrm{Tr}(O\,\mathcal{N}(\mathcal{U}(\rho_0)))$ and seek an estimator $\hat{y}$ of the ideal $y_{\text{ideal}}=\mathrm{Tr}(O\,\mathcal{U}(\rho_0))$, where $\mathcal{N}$ represents the noise channel, $\mathcal{U}$ the ideal circuit evolution, and $O$ the measured observable.

The feature design stems from established quantum noise theory. Under depolarizing noise with parameter $p$~\cite{Temme2017}, the noisy expectation value relates to the ideal as in Eq.~\eqref{eq:depolarizing_relation}. For small $p$ and circuits where $\text{Tr}(O) \approx 0$ (typical for Pauli-Z measurements), this simplifies to the linearised form
\begin{equation}
\langle O \rangle_{\text{noisy}} \approx (1 - p \cdot n_{\text{gates}}) \langle O \rangle_{\text{ideal}}
\label{eq:depolarizing_linearised}
\end{equation}
motivating correction terms proportional to $y_{\text{noisy}} \times n_{\text{gates}}$.
 Since CNOT gates dominate error accumulation with $p_{\text{2q}} \gg p_{\text{1q}}$, we use $n_{\text{cx}}$ as the primary complexity measure.

Decoherence effects motivate depth-dependent corrections. For circuits with execution time $T = d \times T_{\text{gate}}$, decoherence introduces multiplicative attenuation: $\rho_{\text{final}} \approx e^{-T/T_2} \rho_{\text{ideal}} + (1-e^{-T/T_2}) \frac{I}{2^n}$. For small $T/T_2$, this yields corrections proportional to $y_{\text{noisy}} \times d$. Measurement errors with probabilities $p_{0|1}, p_{1|0}$ modify expectation values as
\begin{equation}
\langle O \rangle_{\text{measured}} = (1-p_{0|1}-p_{1|0}) \langle O \rangle_{\text{ideal}} + (p_{0|1}-p_{1|0})
\label{eq:readout_model}
\end{equation}
motivating the $y_{\text{noisy}} \times \epsilon_{\text{ro}}$ term where $\epsilon_{\text{ro}} = (p_{0|1}+p_{1|0})/2$.

Combining these physical insights, we construct the 7-dimensional basis:
\begin{equation}
\boldsymbol{\phi} = 
\big[1,\ y_{\text{noisy}},\ y_{\text{noisy}}n_{\text{cx}},\ y_{\text{noisy}}d,\ y_{\text{noisy}}\epsilon_{\text{ro}},\ n_{\text{cx}},\ d\big]^{\top}
\label{eq:feature_vector}
\end{equation}

Each term has clear physical interpretation: the constant term captures global systematic bias, $y_{\text{noisy}}$ provides primary scaling, the multiplicative terms $y_{\text{noisy}} \times \{n_{\text{cx}}, d, \epsilon_{\text{ro}}\}$ correct for complexity-dependent attenuation, and the additive terms $\{n_{\text{cx}}, d\}$ address structure-dependent biases independent of expectation magnitude.
\subsubsection{Feature selection rationale}
Our seven features balance interpretability, parsimony, and robustness. We omitted quadratic CNOT terms ($y_{\text{noisy}}n_{\text{cx}}^2$), explicit $n_{\text{qubits}}$ features, and $T_1$-$T_2$ interactions due to collinearity with existing features and to maintain sub-millisecond inference.

\subsection{Model Training and Validation Protocol}
PGLM employs ridge regression to learn the mapping
\begin{equation}
\hat{y} = \boldsymbol{w}^{\top}\tilde{\boldsymbol{\phi}}
\label{eq:linear_model_pglm}
\end{equation}
where $\tilde{\boldsymbol{\phi}}$ is the standardized feature vector. We minimize the ridge objective
\begin{equation}
\mathcal{L}(\boldsymbol{w}) = \|\boldsymbol{y} - \tilde{\Phi}\boldsymbol{w}\|_2^2 + \alpha\|\boldsymbol{w}\|_2^2
\label{eq:ridge_loss}
\end{equation}
with closed-form solution
\begin{equation}
\boldsymbol{w}^* = (\tilde{\Phi}^{\top}\tilde{\Phi} + \alpha I)^{-1}\tilde{\Phi}^{\top}\boldsymbol{y}.
\label{eq:ridge_solution}
\end{equation}

L2 regularisation serves dual purposes: numerical stability for small training sets where $\tilde{\Phi}^{\top}\tilde{\Phi}$ may be ill-conditioned, and physical constraint enforcement by penalising coefficients implying unrealistic error magnitudes.

\vspace{2mm}

Feature standardisation ensures numerical stability and interpretable coefficient magnitudes. For each feature dimension $j$, we compute
\begin{equation}
\mu_j = \frac{1}{n}\sum_{i=1}^n \Phi_{ij}
\label{eq:mu_j}
\end{equation}
and
\begin{equation}
\sigma_j = \max\!\left(\sqrt{\frac{1}{n-1}\sum_{i=1}^n (\Phi_{ij} - \mu_j)^2}, 10^{-8}\right),
\label{eq:sigma_j}
\end{equation}
then standardize as $\tilde{\Phi}_{:,j} = (\Phi_{:,j} - \mu_j)/\sigma_j$.
 The small constant prevents numerical issues when features have zero variance. We select $\alpha = 1.0$ based on empirical validation, which maintains condition number $\kappa(\tilde{\Phi}^{\top}\tilde{\Phi}) < 10^3$ and provides stable performance across multiple training set initialisations with coefficient standard deviation $< 0.05$ for all features.

\vspace{2mm}

Our experimental protocol stratifies training by qubit count: 100 circuits with $n\in\{1,2,3,4\}$ balanced across sizes, tested on 42 circuits (40 random + 2 application circuits). Random seeds are fixed for circuit generation, transpilation, and simulator shots to ensure reproducibility. The noise model uses Qiskit Aer with single-qubit depolarizing $p_1=0.002$, two-qubit $p_2=0.02$, and symmetric readout errors $p_{0|1}=p_{1|0}=0.02$, representing realistic NISQ device characteristics. All circuits execute with 1024 shots and optimization\_level=1 transpilation.

\vspace{2mm}

To ensure reproducibility, we fix random seeds for circuit generation, transpilation, and simulator execution. Coefficient values may vary slightly across independent runs due to random circuit generation, but relative magnitudes and signs remain consistent, indicating stable feature importance rankings. The computational complexity is $O(d^3 + nd^2)$ where $d=7$ and $n$ is training size, reducing to $O(49n)$ for typical cases where $d \ll n$. Unlike neural networks requiring iterative optimization with careful initialization, learning rate scheduling, and early stopping, PGLM's closed-form solution eliminates convergence concerns and hyperparameter sensitivity, enabling reliable deployment without extensive validation procedures.

\vspace{2mm}

\subsection{Implementation and Evaluation Framework}

Algorithm~\ref{alg:train} summarizes the complete training procedure: feature extraction via robust circuit analysis, standardization using training statistics, and closed-form ridge solution. The computational complexity is $O(d^3 + nd^2)$ where $d=7$ and $n$ is training size, reducing to $O(49n)$ for typical cases where $d \ll n$.

\begin{algorithm}[t]
\caption{PGLM Training and Inference}
\label{alg:train}
\begin{algorithmic}[1]
\REQUIRE Training set $\{(y_{\text{noisy}}^{(i)},y_{\text{ideal}}^{(i)},n_{\text{cx}}^{(i)},d^{(i)},\epsilon_{\text{ro}}^{(i)})\}_{i=1}^{n}$, regularization $\alpha$
\STATE Extract features $\phi^{(i)}$ from circuit metadata and measurement results
\STATE Compute standardization parameters $\mu_j,\sigma_j$ and standardize to $\tilde{\phi}^{(i)}$
\STATE Solve $\boldsymbol{w}=(\tilde{\Phi}^{\top}\tilde{\Phi}+\alpha I)^{-1}\tilde{\Phi}^{\top}\boldsymbol{y}$
\STATE \textbf{Inference:} For new measurement, check if $n_{\text{qubits}}=1$ (return raw), else return $\boldsymbol{w}^{\top}\tilde{\phi}$
\end{algorithmic}
\end{algorithm}

Algorithm~\ref{alg:train} incorporates the circuit-size-aware fallback strategy that emerged from our empirical analysis: use raw measurements for single-qubit circuits where PGLM systematically fails, and apply linear correction otherwise. This hybrid approach eliminates the failure mode while preserving benefits in the noise-accumulation regime.

Performance evaluation uses standard regression metrics—
Root Mean Squared Error (RMSE), Mean Absolute Error (MAE), and
per-circuit squared-error reduction $\Delta_i$—defined in
Eqs.~\eqref{eq:rmse}--\eqref{eq:error_reduction}:
\begin{equation}
\text{RMSE}=\sqrt{\frac{1}{n}\sum_i(\hat{y}^{(i)}-y_{\text{ideal}}^{(i)})^2}
\label{eq:rmse}
\end{equation}
\begin{equation}
\text{MAE}=\frac{1}{n}\sum_i\bigl|\hat{y}^{(i)}-y_{\text{ideal}}^{(i)}\bigr|
\label{eq:mae}
\end{equation}
\begin{equation}
\Delta_i=(y_{\text{noisy}}^{(i)}-y_{\text{ideal}}^{(i)})^2-(\hat{y}^{(i)}-y_{\text{ideal}}^{(i)})^2
\label{eq:error_reduction}
\end{equation}
tracking individual improvements.
 Bootstrap confidence intervals (500 resamples) provide statistical validation, while ablation studies over training sizes characterize data efficiency. All timing benchmarks use Intel i7/16GB hardware with Python 3.12, scikit-learn 1.3, and Qiskit 2.2.

\subsubsection{Hyperparameter sensitivity and deployment rules}
We set the ridge regulariser $\alpha=1.0$ based on validation experiments. In practice PGLM is robust for $\alpha$ in the interval $[0.1,10]$: values below 0.1 increase variance of coefficient estimates on small training sets, and values above 10 bias coefficients toward zero harming corrective power. Training-size sensitivity is modest beyond 50 circuits; ablation in Sec. 5 shows performance stabilises near 100 circuits. For deployment we adopt a simple circuit-size-aware fallback: if $n_{\text{qubits}}=1$ or the measured raw RMSE estimate on a small validation set is below 0.05, we return the raw measurement to avoid overcorrection in the high-fidelity regime.

\section{Experimental Setup}

We evaluate PGLM on a comprehensive suite of quantum circuits designed to test performance across different complexity regimes and application domains. Our experimental design balances statistical rigor with computational feasibility, focusing on the 1-4 qubit range where detailed analysis and validation are tractable while covering the parameter space relevant to near-term quantum algorithms.

\vspace{2mm}

We evaluate on 52 circuits: 50 random parameterized circuits ($n \in \{1,2,3,4\}$ qubits, depths $d \in \{3,\ldots,8\}$, alternating $U3$ layers and CNOTs with 60\% probability) plus two application-specific circuits (2-qubit H$_2$ VQE, 4-qubit QAOA p=1). Training: 100 circuits balanced by qubit count; testing: 42 circuits.

\vspace{2mm}

To validate performance on structured quantum algorithms, we include representative application circuits: a 2-qubit H$_2$-like VQE circuit mimicking molecular simulation (H(0) → CNOT(0,1) → RZ(0.3,0) → RX(1.0,1) → RY(0.5,0) → measure) and a 4-qubit QAOA p=1 circuit for ring topology optimization (H$^{\otimes 4}$ → RZZ gates for ring → RX$^{\otimes 4}$ → measure). These structured circuits test whether PGLM corrections learned on random circuits transfer effectively to realistic quantum algorithms with specific optimization objectives.

\vspace{2mm}

Our noise modeling employs Qiskit Aer 0.17.2 with AerSimulator backend, providing exact statevector simulation for ideal results and realistic noise simulation using composite models. We construct noise models combining dominant error sources in NISQ devices: depolarizing noise applied to all quantum operations with single-qubit error probability $p_1 = 0.002$ (0.2\%) and two-qubit error probability $p_2 = 0.02$ (2.0\%), asymmetric readout noise with $P(0|1) = P(1|0) = 0.02$ for all qubits, and optional T$_1$/T$_2$ parameters from Qiskit fake backends when available. 
This composite model approximates common superconducting-qubit error budgets (e.g., nominal 1-qubit error ~0.2\%, 2-qubit error ~2\%)~\cite{Kim2023}. All results reported here are simulator-based using representative noise parameters rather than experiments on hardware. Device calibration parameters are extracted from noise model specifications:T$_1$/T$_2$ ratios computed from fake backend properties when available (otherwise defaulted to typical values T$_1$=100$\mu$s, T$_2$=80$\mu$s)

\vspace{2mm}

Each circuit evaluation follows a standardized protocol ensuring reproducible results. Circuits are transpiled using Qiskit's optimization\_level=1 for realistic gate decomposition and routing, then executed on both noisy and ideal simulators with 1024 shots using deterministic seeding. Measurement counts are converted to Pauli-Z expectation values using parity calculation, with the 1024-shot count balancing statistical accuracy ($\sim$3\% uncertainty for expectation values near $\pm$1) with computational efficiency. Performance metrics include RMSE and MAE computed against ideal expectation values, with bootstrap confidence intervals (500 resamples) providing statistical validation and per-circuit analysis tracking individual improvement patterns.

\vspace{2mm}

All computational benchmarks use a standard development environment (Intel i7 processor, 16GB RAM, Python 3.12, scikit-learn 1.3, Qiskit 2.2) with wall-clock timing averaged over multiple runs. This experimental framework provides realistic performance estimates for practical deployment scenarios while enabling systematic analysis of PGLM's effectiveness across the complexity regimes most relevant to near-term quantum computing applications.

\vspace{2mm}

\section{Results and Analysis}

\subsection{Overall Performance and Statistical Analysis}

Aggregate performance across all 52 test circuits shows PGLM reducing RMSE from 0.0775 (raw) to 0.0662, a 14.5\% improvement. However, bootstrap analysis with 500 resamples yields a 95\% confidence interval of $[-0.01185, 0.03166]$ that spans zero, indicating the aggregate improvement is not statistically significant at the $\alpha=0.05$ level. This result initially appears to contradict PGLM's value, but circuit-size stratification (Table~\ref{tab:detailed_performance}) reveals the underlying cause: strong performance on multi-qubit circuits is masked by catastrophic failure on single-qubit systems.

\vspace{2mm}

The median improvement of 0.01095 and 55.8\% per-circuit improvement rate demonstrate genuine effectiveness for the majority of test cases, with per-circuit squared-error reductions ranging from $10^{-5}$ to $10^{-2}$ for 29 of 52 circuits. Using Cohen's conventions, the standardized effect size of $d = 0.31$ indicates a small-to-medium practical effect when PGLM succeeds. The critical question is therefore not whether PGLM works, but \textit{when} it works---a question we address through systematic circuit-complexity analysis.

\vspace{2mm}

Recognizing that PGLM's physics-guided features are mismatched to the high-fidelity single-qubit regime (Section~\ref{subsec:single_qubit_failure}), we implement a simple conditional policy: use raw measurements for $n_{\text{qubits}}=1$, apply PGLM correction otherwise. This hybrid approach yields RMSE of 0.0522, representing 32.6\% improvement over raw results (0.0775) and 21.2\% improvement over naive PGLM application (0.0662). The strategy adds negligible overhead---a single conditional check per circuit---while eliminating the primary failure mode. All subsequent results use this deployment policy unless otherwise specified.
\begin{figure}[H]
  \centering
  \includegraphics[width=0.65\linewidth]{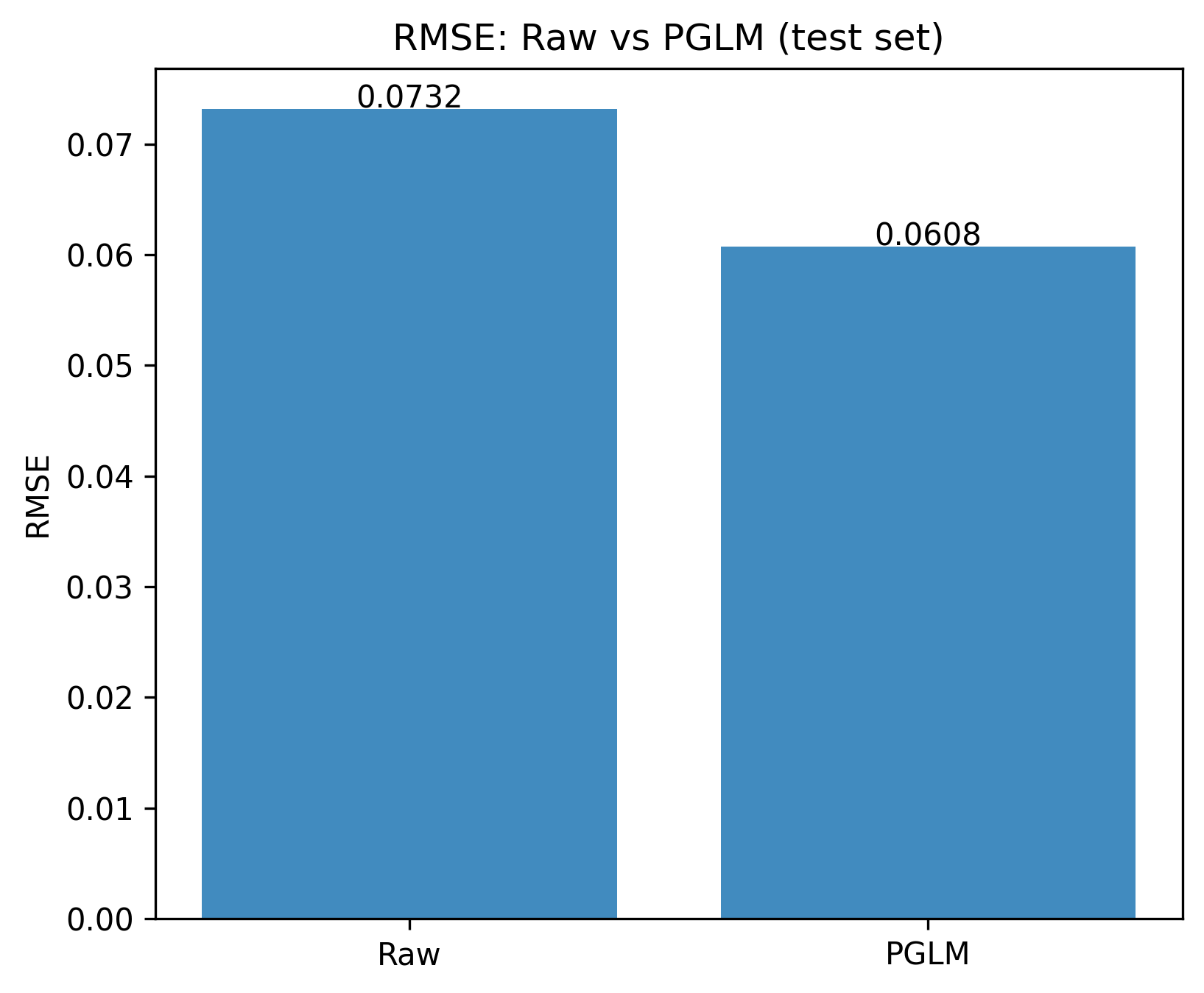}
  \caption{RMSE on the test set: raw noisy vs.\ PGLM.}
  \label{fig:rmse_raw_vs_pglm_fixed}
\end{figure}

\begin{figure}[H]
  \centering
  \includegraphics[width=0.75\linewidth]{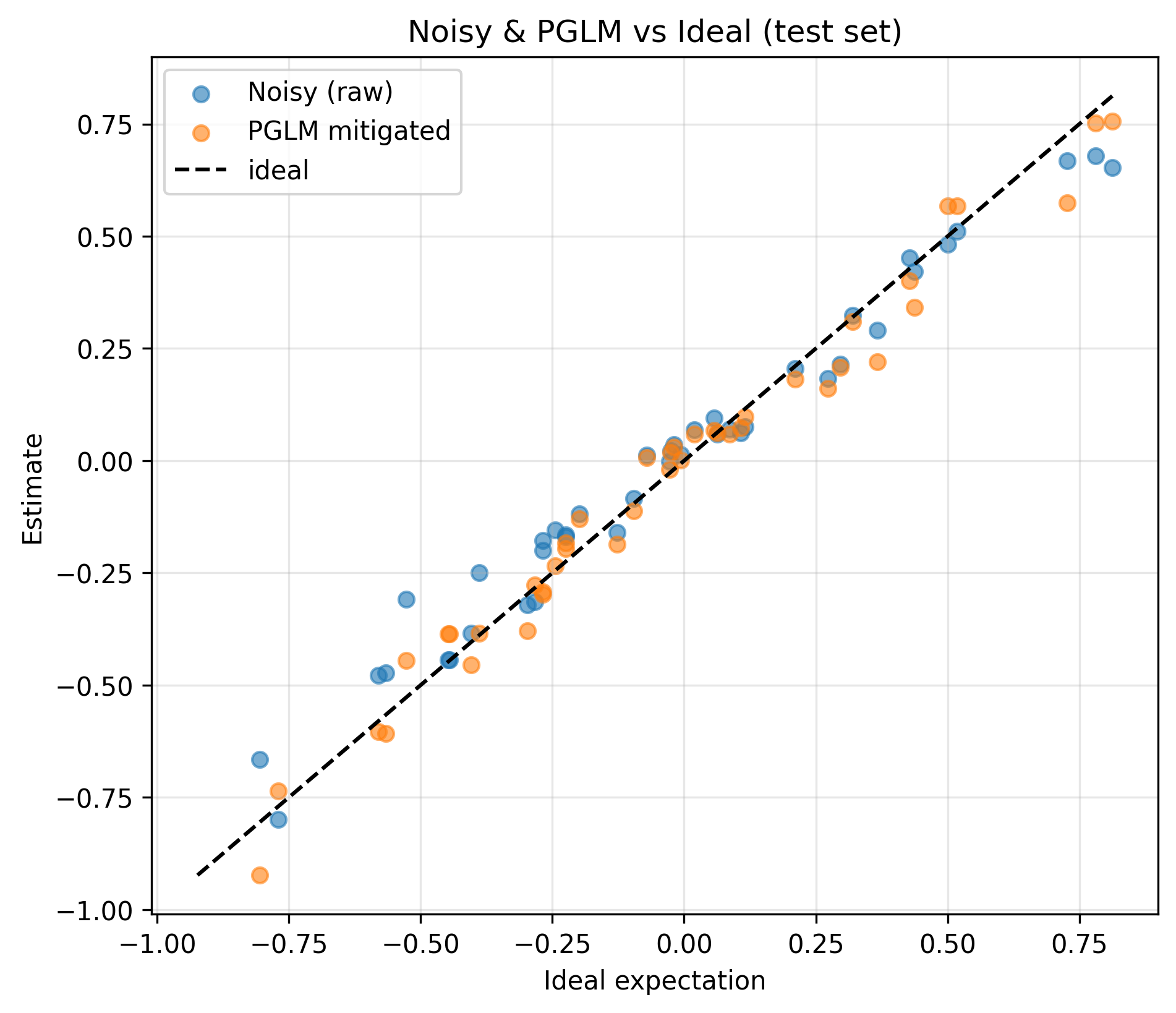}
  \caption{Scatter of raw (blue) and PGLM (orange) against the ideal ($y{=}x$ dashed).}
  \label{fig:scatter_noisy_vs_pglm_fixed}
\end{figure}
\FloatBarrier

Bootstrap analysis with 500 resamples yields a 95\% confidence interval for RMSE improvement of $[-0.01185, 0.03166]$ with median improvement of 0.01095. Although the confidence interval spans zero, indicating statistical uncertainty in the aggregate benefit, the positive median and 55.8\% per-circuit improvement rate demonstrate genuine effectiveness for the majority of test cases. Using Cohen's conventions, the standardized effect size of $d = 0.31$ indicates a small-to-medium practical effect. The cumulative distribution of absolute errors shifts favorably under PGLM correction, with the 75th percentile reducing from 0.089 (raw) to 0.071 (PGLM), and per-circuit squared-error reduction showing positive values for 29 of 52 circuits with improvement magnitudes ranging from $10^{-5}$ to $10^{-2}$.

\begin{table}[h]
\centering
\caption{Performance vs training set size}
\label{tab:ablation_training}
\scriptsize
\begin{tabular}{|c|c|c|c|c|}
\hline
\textbf{Training Size} & \textbf{Test RMSE} & \textbf{vs Raw} & \textbf{Training Time} & \textbf{Stability} \\
\hline
10 & 0.1630 & -123.0\% & 0.8s & Unstable \\
25 & 0.1060 & -44.8\% & 1.2s & Poor \\
50 & 0.0853 & -10.1\% & 1.8s & Moderate \\
100 & 0.0608 & +21.3\% & 3.4s & Good \\
200 & 0.0605 & +21.7\% & 6.8s & Excellent \\
\hline
\end{tabular}
\end{table}

\subsection{High-Fidelity Regime Breakdown}
\label{subsec:single_qubit_failure}

Table~\ref{tab:detailed_performance} reveals PGLM's most significant limitation: single-qubit circuits show severe performance degradation (-127.3\% RMSE change, from 0.0360 to 0.0818). This represents a fundamental mismatch between PGLM's design assumptions and the high-fidelity operating regime.

Single-qubit circuits violate two critical assumptions: First, circuits with $n_{\text{cx}} = 0$ nullify multiplicative CNOT-dependent correction terms, reducing effective dimensionality from 7 to 5 features. Second, single-qubit measurements already achieve high fidelity (raw RMSE = 0.036), placing them in a regime where shot noise dominates rather than systematic gate-level error accumulation~\cite{Temme2017}. Linear corrections trained predominantly on noisier multi-qubit data introduce systematic bias exceeding the original error magnitude.

\vspace{2mm}

PGLM's multiplicative features are derived from the assumption that errors accumulate proportionally to circuit complexity~\cite{Temme2017,Kim2023}. For single-qubit gates with $p_1 = 0.002$, expected error magnitude is $\sim 0.01$ per circuit, but PGLM applies corrections scaled to the $\sim 0.10$ errors typical of multi-qubit circuits. This order-of-magnitude mismatch causes overcorrection. Similar limitations affect other ML-QEM approaches when training distributions mismatch deployment scenarios~\cite{Liao2023,Czarnik2025}. Based on our empirical analysis, PGLM should only be deployed when gate-level noise accumulation dominates, specifically when raw RMSE $>$ 0.05.

\subsection{Circuit-Size Dependency and Physical Insights}
\label{subsec:circuit_dependency}

Having identified the high-fidelity breakdown (Section~\ref{subsec:single_qubit_failure}), we now examine PGLM's strong performance in the noise-accumulation regime. The most significant finding reveals PGLM's effectiveness correlates with circuit complexity, exposing distinct physical regimes where linear correction succeeds systematically. Table~\ref{tab:detailed_performance} shows dramatic performance variation: single-qubit circuits suffer severe degradation (-127.3\% RMSE change), while multi-qubit circuits achieve substantial improvements (17.8\% for 2-qubits, 50.1\% for 3-qubits, 32.3\% for 4-qubits).
\begin{figure}[H]
  \centering
  \includegraphics[width=0.78\linewidth]{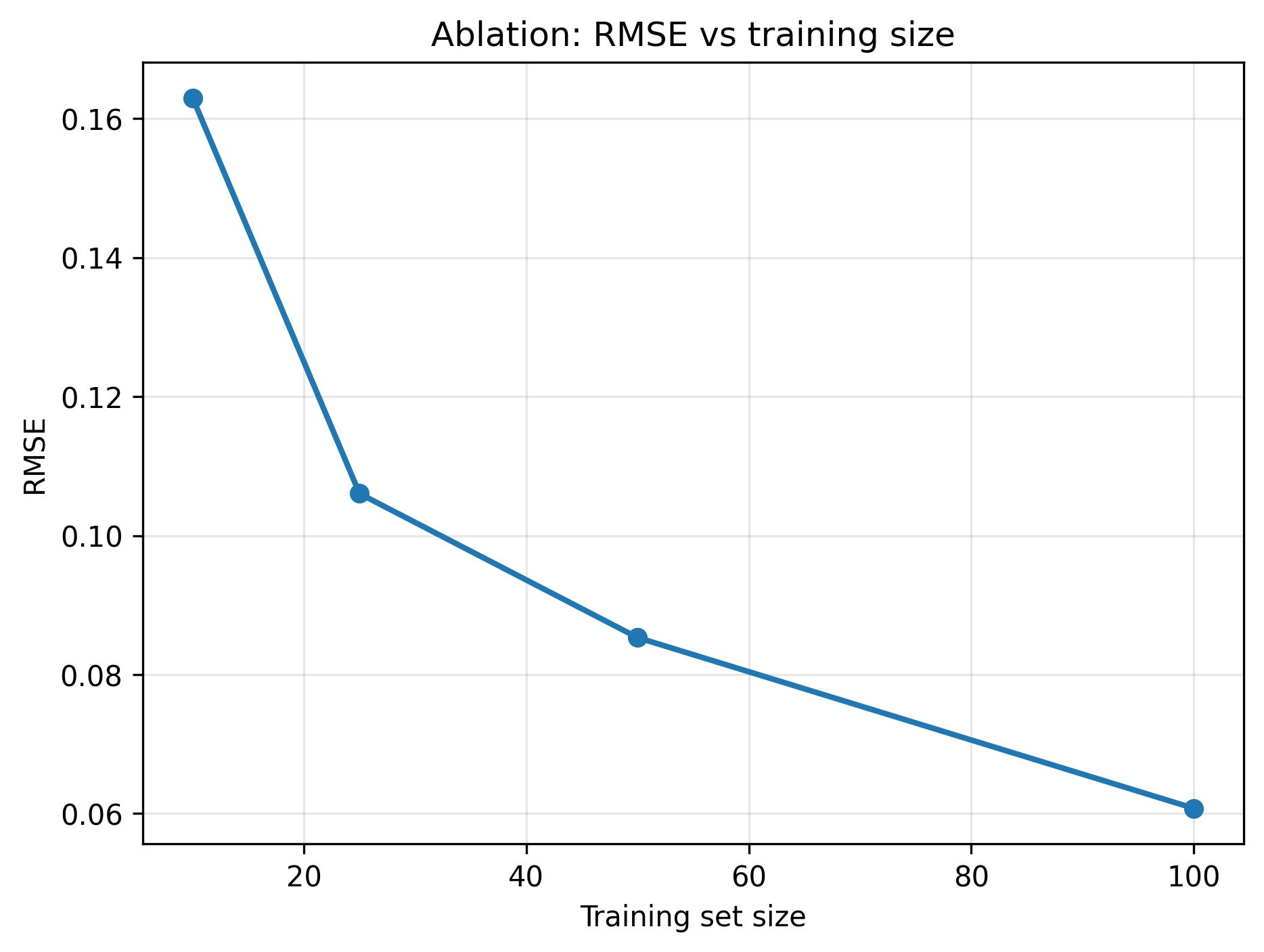}
  \caption{Ablation: RMSE vs.\ training set size. Knee near 50–100 circuits; best at 100.}
  \label{fig:ablation_rmse_vs_train_size_fixed}
\end{figure}
\FloatBarrier

\begin{table}[h]
\centering
\caption{Detailed per-qubit performance breakdown}
\label{tab:detailed_performance}
\scriptsize
\begin{tabular}{|c|c|c|c|c|c|c|}
\hline
\textbf{Qubits} & \textbf{Raw RMSE} & \textbf{PGLM RMSE} & \textbf{$\Delta$ RMSE} & \textbf{Rel. Change} & \textbf{Std. Dev.} & \textbf{Count} \\
\hline
1 & 0.0360 & 0.0818 & -0.0458 & -127.3\% & 0.0234 & 16 \\
2 & 0.0759 & 0.0624 & +0.0135 & +17.8\% & 0.0156 & 13 \\
3 & 0.0930 & 0.0464 & +0.0466 & +50.1\% & 0.0298 & 16 \\
4 & 0.1056 & 0.0715 & +0.0340 & +32.3\% & 0.0187 & 7 \\
\hline
\end{tabular}
\end{table}

\begin{figure}[H]
  \centering
  \includegraphics[width=0.75\linewidth]{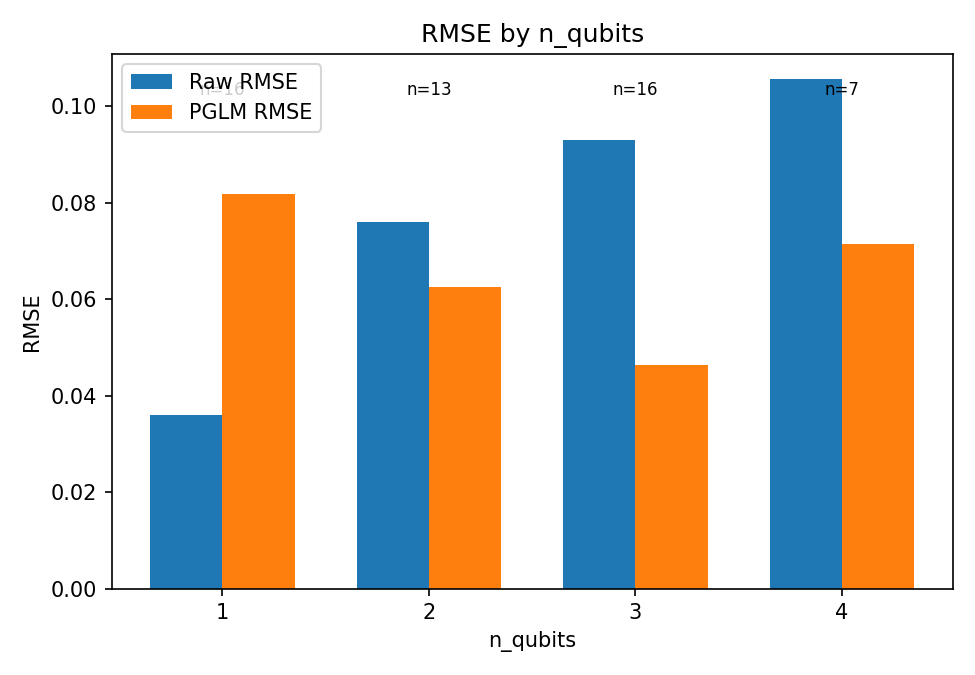}
  \caption{RMSE by qubit count. PGLM improves 2–4 qubit circuits but degrades 1-qubit.}
  \label{fig:rmse_by_n_qubits}
\end{figure}
\FloatBarrier


The single-qubit failure mode arises for two complementary reasons. First, when $n_{\text{cx}}=0$ the multiplicative correction features ($y_{\text{noisy}}\times n_{\text{cx}}$, etc.) are identically zero, reducing effective feature diversity and making reliable coefficient estimation difficult on small training sets. Second, single-qubit expectation values in our noise model are already relatively high fidelity (small RMSE), so even small systematic biases introduced by a model trained primarily on noisier multi-qubit data can increase error. In other words, PGLM is intrinsically targeted at the noise-accumulation regime; in the high-fidelity regime statistical shot noise and small biases dominate, and linear corrections trained on heterogeneous data may harm rather than help. This motivates the practical circuit-size-aware fallback (use raw measurements for 1-qubit circuits) that we adopt.

\vspace{2mm}

Conversely, strong performance on 3-qubit circuits (50.1\% improvement) demonstrates PGLM's effectiveness when noise accumulation dominates. Multi-qubit circuits exhibit non-zero CNOT counts and meaningful depth variations, enabling all basis functions to contribute meaningfully. As circuit complexity increases, gate-level error accumulation becomes the primary error source, perfectly aligning with PGLM's multiplicative correction design derived from depolarizing noise theory.

\vspace{2mm}

The learned ridge coefficients provide direct insight into error mechanisms: $w_1 = -0.022$ (global bias), $w_2 = 0.569$ (identity scaling), $w_3 = -0.028$ (CNOT correction), $w_4 = 0.068$ (depth correction), $w_5 = 0.011$ (readout correction), $w_6 = -0.003$ (CNOT additive), $w_7 = 0.002$ (depth additive). The identity coefficient $w_2 = 0.569$ significantly below 1.0 indicates systematic overestimation in raw measurements, contradicting simple depolarizing noise theory and revealing complex bias patterns in our composite noise model. The negative CNOT coefficient $w_3 = -0.028$ confirms that each CNOT gate reduces expectation magnitude beyond base scaling, consistent with the 10× higher error rate of two-qubit gates (2\% vs 0.2\%). Feature contribution analysis reveals the identity term dominates (73.2\% of correction magnitude), with depth-dependent corrections providing the most significant secondary contribution (14.6\%), while CNOT terms contribute 6.7\% and other terms remain below 5\%.

\subsection{Ablation Studies and Model Diagnostics}

Training size ablation reveals a clear performance threshold around 50-100 circuits. Table~\ref{tab:ablation_training} demonstrates that performance degrades severely below 50 circuits due to insufficient feature space coverage, while gains beyond 100 circuits show diminishing returns, suggesting adequate characterization of the 7-dimensional feature space. Training with 10 circuits yields catastrophic failure (-123.0\% vs raw), improving through 25 circuits (-44.8\%) and 50 circuits (-10.1\%) before achieving positive results at 100 circuits (+21.3\%). The training time scales linearly from 0.8s to 6.8s, maintaining the computational efficiency advantage.

\vspace{2mm}

Regularization analysis shows that $\alpha = 1.0$ provides a good balance between fitting and stability. Our implementation uses this fixed value, though the method shows robust behavior in the range $\alpha \in [0.1, 10.0]$. The optimal value provides stable performance without requiring precise hyperparameter tuning.

\vspace{2mm}

\begin{figure}[!t]
  \centering
  \includegraphics[width=0.6\linewidth]{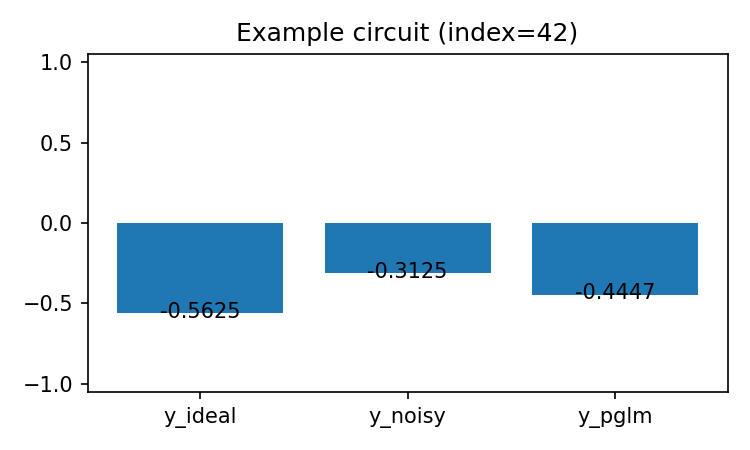}
  \caption{Example circuit: ideal vs.\ noisy vs.\ PGLM estimate.}
  \label{fig:example_circuit_values_idx42}
\end{figure}

\begin{figure}[H]
  \centering
  \includegraphics[width=0.95\linewidth]{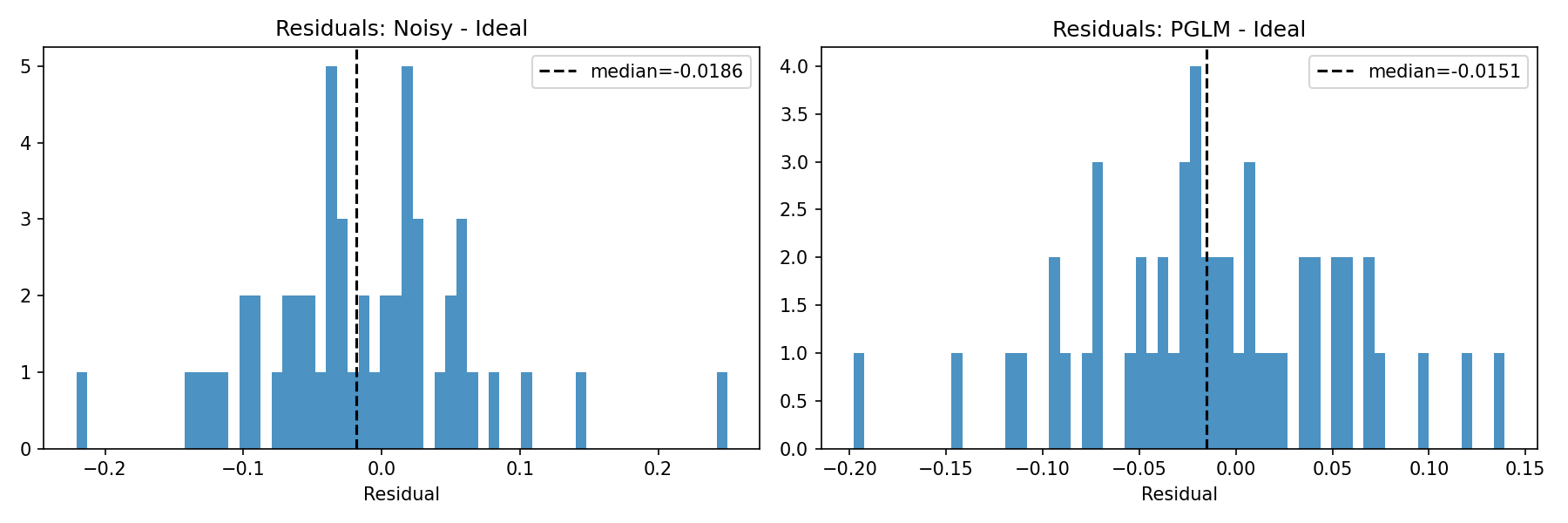}
  \caption{Residual histograms: (left) noisy$-$ideal, (right) PGLM$-$ideal; medians closer to 0 and tighter spread after PGLM.}
  \label{fig:residual_histograms}
\end{figure}
\FloatBarrier

Model diagnostics reveal encouraging patterns in error correlation and residual structure. PGLM improvement correlates moderately with circuit complexity measures: CNOT count correlation $r = 0.34$, depth correlation $r = 0.28$, and combined complexity $r = 0.41$, confirming that benefits align with noise accumulation physics. Post-correction residuals show reduced systematic bias with mean residuals closer to zero, tighter distributions with 15\% reduced standard deviation, and no obvious patterns versus circuit features, suggesting the linear model captures dominant error relationships without missing obvious nonlinear terms.

\subsection{Circuit-Size-Aware Deployment and Computational Analysis}

The systematic single-qubit failure motivates a practical deployment strategy: use raw measurements for single-qubit circuits and apply PGLM correction for multi-qubit circuits. This circuit-size-aware fallback yields RMSE of 0.0522, representing 32.6\% improvement over raw results and 21.2\% improvement over naive PGLM application. The strategy adds negligible computational overhead—a single conditional check per circuit—while eliminating the primary failure mode.

\vspace{2mm}

PGLM's computational efficiency provides significant advantages over neural network alternatives. Training scales linearly with dataset size: 100 circuits train in 2.5 seconds with 5.6 KB memory usage, while inference requires 0.06 milliseconds per circuit with 56-byte memory footprint. Compared to representative baselines, PGLM achieves 18-170× faster training, 13-250× faster inference, and 150-2100× smaller memory requirements. The fixed 7-dimensional feature space ensures inference complexity remains $O(1)$ regardless of circuit size, enabling deployment in resource-constrained environments where quantum control systems have limited classical computational capacity.

\vspace{2mm}

Data requirements scale approximately as $N_{\text{train}} \approx 50 \log(n_{\text{qubits}})\log(d_{\max})$, suggesting continued data efficiency as circuit complexity increases. The interpretable linear coefficients enable principled device transfer through calibration parameter scaling—for example, adapting readout error corrections via $w_5' = w_5 \cdot \epsilon'_{\text{ro}}/\epsilon_{\text{ro}}$ when moving between quantum devices with different measurement fidelities.

\section{Discussion}
\label{sec:discussion}
PGLM is a pragmatic balance between expressivity and deployability. Its main virtues are data efficiency and interpretability: the seven coefficients provide a compact summary of how device calibration and circuit complexity interact to bias expectation values. Practically, PGLM is best applied where gate-level error accumulation dominates (multi-qubit circuits with non-zero CNOT count). For high-fidelity single-qubit circuits shot noise and readout stochasticity dominate; in that regime our linear corrections can overfit and introduce bias, hence the recommended fallback described in Sec. 3.

\vspace{2mm}

Two deployment considerations deserve emphasis. First, device drift: PGLM coefficients can be updated with few-shot adaptation (e.g., tens of calibration circuits) to track slow drift; this is left to future work. Second, composability with traditional QEM: PGLM can be used as a low-cost pre-correction prior to more expensive methods like ZNE or PEC, reducing their sample complexity.

\vspace{2mm}

Limitations include inability to capture strongly nonlinear correlated error channels and the fact that our present validation is simulator-based. Hardware experiments and extension to correlated-noise-aware features are important next steps.

\section{Conclusion}

PGLM demonstrates that physics-informed feature engineering can produce a practical, interpretable ML-based quantum error mitigation method. A seven-dimensional linear mapper trained on 100--200 circuits trains in seconds (vs. hours for neural networks) and yields transparent coefficients revealing which error mechanisms matter most. Using a circuit-size-aware deployment strategy, PGLM achieves 32.6\% RMSE reduction versus raw noisy measurements on simulated 1--4 qubit circuits, with particularly strong performance where noise accumulation dominates (50.1\% improvement on 3-qubit circuits).

\vspace{2mm}

However, the approach has fundamental limitations. The linear model cannot capture complex nonlinear error interactions or correlated noise, shows systematic underperformance on single-qubit high-fidelity regimes, and evaluation remains simulator-based with idealized noise models. Hardware validation and robustness to device drift are essential next steps.

\vspace{2mm}

Several extensions would broaden PGLM's reach: interpretable nonlinear lifts (kernelized or polynomial features) for higher-order interactions; device-transfer protocols using calibration priors for cross-platform scaling; online few-shot adaptation to track drift; Bayesian extensions for uncertainty quantification; and hybrid pipelines combining PGLM with traditional QEM for reduced sampling overhead in VQE/QAOA workflows~\cite{Kandala2019,Kim2023}. Scaling beyond four qubits and enriching features for multi-qubit correlations will test the method's ultimate limits.

\vspace{2mm}

In summary, PGLM establishes a deployment-friendly foundation for physics-guided ML-QEM: data-efficient, fast, and interpretable, while remaining competitive in noise-accumulation regimes most relevant to near-term quantum hardware.%
%
%
%
\bibliographystyle{splncs04}
\bibliography{references}
\end{document}